\documentclass[12pt]{article}
\usepackage{amsmath,amssymb,amsthm,cite,enumitem,mathtools,xcolor}

\usepackage[english]{babel}
\usepackage{graphicx}
\usepackage[T1]{fontenc}
\usepackage[cp1250]{inputenc}
\usepackage{appendix}
\usepackage{verbatim}
\usepackage{color}
\usepackage{tikz}
\usetikzlibrary{calc}
\usetikzlibrary{decorations.pathmorphing}
\usetikzlibrary{decorations.shapes}
\usetikzlibrary{arrows}
\usetikzlibrary{decorations.pathreplacing}
\usepackage{amsmath}
\usepackage{amsfonts}
\usepackage{ragged2e}
\allowdisplaybreaks

\newcommand\e{\mathrm{e}}
\newcommand\ren{\mathrm{ren}}

\newcommand\al{\alpha}
\newcommand\be{\beta}
\newcommand\ga{\gamma}
\newcommand\Ga{\Gamma}
\newcommand\de{\delta}

\newcommand\eps{\epsilon}
\newcommand\ka{\kappa}
\newcommand\la{\lambda}
\newcommand\La{\Lambda}
\newcommand\si{\sigma}

\newcommand\cA{{\mathcal A}}

\newcommand\cC{{\mathcal C}}
\newcommand\cD{{\mathcal D}}

\newcommand\cJ{{\mathcal J}}

\newcommand\cO{{\mathcal O}}

\newcommand\cS{{\mathcal S}}

\newcommand\one{{\bf{1}}}

\newcommand\rd{{\rm d}}
\newcommand\re{{\rm e}}
\newcommand\ri{{\rm i}}

\newcommand\beq{\begin{equation}}
\newcommand\eeq{\end{equation}}
\newcommand\beaX{\begin{eqnarray*}}
\newcommand\eeaX{\end{eqnarray*}}
\newcommand\bea{\begin{align}}
\newcommand\eea{\end{align}}
\newcommand\btp{\begin{tikzpicture}}
\newcommand\etp{\end{tikzpicture}}
\newcommand\nn{\notag}

\newcommand\rarr{\rightarrow}

\newcommand\pd{\partial} 
\newcommand\tr{\text{tr}}
\newcommand\Tr{\text{Tr}}

\newcommand\pslash{p \hspace{-1.0ex}\slash}

\newcommand\kslash{k \hspace{-1.1ex}\slash}

\newcommand\qslash{q \hspace{-1.3ex}\slash}

\newtheorem{theoreme}{Theorem }[section]

\newtheorem{lemma}[theoreme]{Lemma}
\newtheorem{definition}[theoreme]{D
efinition}
\newtheorem{corollary}[theoreme]{Corollary}

\newtheorem{example}[theoreme]{Example}

\newcommand{\bes}{\begin{subequations}}
\newcommand{\ees}{\end{subequations}}
\newcommand\bel{\begin{lemma}}
\newcommand\eel{\end{lemma}}
\newcommand\bet{\begin{theoreme}}
\newcommand\eet{\end{theoreme}}
\newcommand\bex{\begin{example}}
\newcommand\eex{\end{example}}
\newcommand\bed{\begin{definition}}
\newcommand\eed{\end{definition}}
\newcommand\bec{\begin{corollary}}
\newcommand\eec{\end{corollary}}

\begin{document}

\title{Axial anomaly in the presence \\of arbitrary spinor interactions}

\author{Jan Derezi\'{n}ski and Adam Latosi\'{n}ski\\
Department of Mathematical Methods in Physics, Faculty of Physics, \protect\\
University of Warsaw, Pasteura 5, 02-093 Warszawa, Poland, \protect\\ 
email: jan.derezinski@fuw.edu.pl, adam.latosinski@fuw.edu.pl}

\date{\today}

\maketitle

\begin{abstract}
We consider $N$ Dirac fermions on a 4-dimensional Euclidean space with a quadratic
interaction given by arbitrary external Clifford-valued fields. The
 divergence of the axial current satisfies on the classical level
 a relation that
 is violated after quantization.
Using the Pauli-Villars method to regularize the fields, we find the
conditions that guarantee  the finiteness of the anomaly. We also find this 
anomaly. 
Our result generalizes the 
well-known computation of axial anomaly of Dirac fermions interacting with an
external Yang-Mills field.
  \end{abstract}

\section{Introduction}

We shall consider a set of  classical anti-commuting Dirac spinor fields $\psi_I,\bar\psi_I$,
$I=1,\dots,N$,  in a
4-dimensional Euclidean space (with positive signature). We also
introduce
a quadratic Hermitian  action for these fields of the form
\beq \cS[\psi] = \int \rd^4x\, \sum_{IJ} \bar\psi_I(x)\big ((\gamma^\mu\pd_\mu  + m)\de^{IJ} + \Phi^{IJ}(x) \big)\psi_J(x).\label{lagran}\eeq
Thus all fields have the same mass $m$ and are coupled
 to external Clifford-algebra-valued fields $\Phi^{IJ}(x)$.

Conserved or approximately conserved currents play an important role
in QFT. For instance, the vector current
\beq \cJ^\mu(x):=\sum_I\bar\psi_I(x)\gamma^\mu\psi_I(x)\eeq satisfies formally
\beq \partial_\mu \cJ^\mu(x)=0.\label{clas}\eeq
The relation \eqref{clas} is true also inside correlation functions:
\beq\langle \partial_\mu
\cJ^\mu(x)\rangle=0.\label{clas0}\eeq

To be precise,  \eqref{clas} is not fully rigorous, since it involves 
putting two fields at coinciding points. The relation  \eqref{clas0}
is exact, since we only consider such regularizations that keep this condition true.

Let us also introduce the following currents:
\beq \cJ^{5\mu}(x) = \sum_I\bar\psi_I(x) \ga^5\ga^\mu \psi_I(x), \qquad  \cJ^{5}(x) = \sum_I \bar\psi_I(x) \ga^5 \psi_I(x). \label{curr0}\eeq
If the Dirac field is coupled  only to an external 
 Yang-Mills field,
 that is,
 the field $\Phi$ has just one component $\Phi(x) =  \ri
 A_\mu(x) \ga^\mu(x)$,  and fields $\psi$ satisfy the classical equations of motion, then  formally on the classical level
 \eqref{curr0} satisfy the equality 
\beq \pd_\mu \cJ^{5\mu}(x) + 2 m \cJ^{5}(x) = 0.\label{clas1}\eeq
However on the quantum level, after appropriate renormalization, we get
\beq \langle\pd_\mu \cJ^{5\mu}(x)\rangle_\ren + 2m  \langle\cJ^{5}(x)\rangle_\ren  = \frac{1}{16\pi^2} \Tr\big(\eps^{\mu\nu\rho\si} A_{\mu\nu}(x) A_{\rho\si}(x)\big) =: \cA(x).\eeq
where $A_{\mu\nu} = \pd_\mu A_\nu - \pd_\nu A_\mu + \ri[A_\mu, A_\nu]$ and Tr denotes the trace over the space enumerating different fields $\psi_I$. $\cA(x)$ is called the axial anomaly. This is described in essentially
every modern textbook on Quantum Field Theory such as  \cite{Sr,We},
see also \cite{FuSu,Be} for a more specialized treatment.

For more general external fields the  classical relation analogous to \eqref{clas1}
is
\beq \pd_\mu \cJ^{5\mu}(x) + 2m \cJ^{5}(x) + \sum_{IJ} \bar\psi_I(x) \big(\ga^5\Phi^{IJ}(x) +\Phi^{IJ}(x)\ga^5\big)\psi_J(x) = 0.\eeq
which holds if the fields $\psi$ satisfy the classical equations of motion.
Since this is a generalization of the simpler case, we still expect an anomaly to be present on the quantum level:
\begin{align}\notag \langle\pd_\mu \cJ^{5\mu}(x)\rangle_\ren + 2m \langle\cJ^{5}(x) \rangle_\ren&\\+ \sum_{IJ} \langle\bar\psi_I(x) \big(\ga^5\Phi^{IJ}(x) +\Phi^{IJ}(x)\ga^5\big)\psi_J(x)\rangle_\ren& = :\cA(x). \label{anomaly-definition0} \end{align}

As in the case of the vector current, $\cJ^{5\mu}(x)$ and the equations
\eqref{anomaly-definition0} are problematic, because they involve two fields
at coinciding points. In order to give sense to them we
shall use the Pauli-Villars regularization. We will  show that with this
regularization   the renormalized vector current $\cJ^\mu(x)$ is
conserved. We will find conditions on the external fields
$\Phi^{IJ}(x)$ which guarantee that the axial anomaly is finite. These
conditions say that certain local  polynomials in external fields and
their derivatives  vanish.  There  are three such
  conditions: one involves a polynomial of degree 1, another of degree
  2 and the third of degree 3.  We
will compute the anomaly---again, given by a local polynomial in
 external fields.   This is a rather complicated
   polynomial of degree four. One  should note that there are no
   anomalies of degree five and more.

 Lagrangians of the form \eqref{lagran} may appear in phenomenological
 description of various physical systems. Our results show that the usual
 approximate conservation of the axial current  can often be
 generalized to a more general setting.

\section{Euclidean Dirac field}

Consider the Clifford algebra generated by matrices $\ga_\mu$ such that
\beq
\{\ga_\mu, \ga_\nu\} =  \ga_\mu\ga_\nu+\ga_\nu\ga_\mu=2g_{\mu\nu},\qquad\ga_\mu^\dagger=\ga_\mu,\quad \mu,\nu=1,\dots,4.\label{cliff}\eeq
where $g_{\mu\nu}$ is the Euclidean metric tensor $g={\rm
  diag}(1,1,1,1)$. We'll use the conventions $\ga_5=\ga_1\ga_2\ga_3\ga_4$, $\gamma_{\mu\nu}:=\frac{1}{2}(\ga_\mu\ga_\nu-\ga_\nu\ga_\mu)$,  and $\bar\psi_I =
(\psi_I)^\dag \ga_5$. Thus \eqref{cliff} can be extended to $\mu,\nu=5$, and
$\ga_{\mu\nu}=-\ga_{\mu\nu}^\dagger=-\ga_{\nu\mu}$ for $\mu,\nu=1,\dots,5 $. Still, whenever contraction of such indices appears in the text, the summation it denotes is only over the range $\{1\dots 4\}$.

The action \eqref{lagran} can be written as
\beq \cS[\psi] = \int \rd^4x\, \sum_{IJ} \bar\psi_I(x) D(m,\Phi)^{IJ} \psi_J(x), \eeq
where
\beq D(m,\Phi)^{IJ}  =  (\gamma^\mu\pd_\mu  + m)\de^{IJ} + \Phi^{IJ} = D_0(m) \de^{IJ} + \Phi^{IJ}.\eeq
With the chosen conventions the free part of the action is Hermitian. For the whole action to be Hermitian, we need
\beq (\Phi^{IJ})^\dag = \ga^5 \Phi^{JI} \ga^5.\eeq
Note that with this assumption $\gamma^5D(m,\Phi)$ is a Hermitian operator.

 It can be noted that  a rotation of fields $\psi^I(x)
  \rarr U^{IJ}(x)\psi^J(x)$, with $U$ being unitary matrices,  is
  equivalent to an apprioprate transformation of the external fields $\Phi$: 
\beq \cS[U\psi] = \int \rd^4x\, \sum_{IJ} \bar\psi_I(x) \big(U^{-1}D(m,\Phi)U\big)^{IJ} \psi_J(x), \eeq
\beq U^{-1}D(m,\Phi)U = D\big(m, U^{-1}\Phi U  + \ga^\mu U^{-1}(\pd_\mu U)\big) \label{gauge-transformation}\eeq

Given a basis of the Clifford algebra, the  external fields $\Phi^{IJ}$ can be decomposed in this basis. We will use a specific basis
\beq \Ga^a \in (\one, \ri\ga^\mu, \ri\ga^{\mu\nu}, \ga^\mu\ga^5, \ga^5) \eeq
and introduce varying symbols for different components of $\Phi^{IJ}$:
\beq \Phi^{IJ} = \Phi^{IJ}_a\Gamma^a = \kappa^{IJ} \one + \ri A^{IJ}_\mu \ga^\mu + \ri B^{IJ}_{\mu\nu} \ga^{\mu\nu} + C^{IJ}_\mu \ga^\mu\ga^5 + \la^{IJ} \ga^5 \eeq
 In some formulas we will also use $\tilde{B}_{\mu\nu} = \frac12 \eps_{\mu\nu\rho\si} B^{\rho\si}$.
With this choice, $(\Gamma^a)^\dag =
\ga^5\Gamma^a\gamma^5$, which further implies that the interaction
part of the action $\cS$ is Hermitian iff every component field
$\Phi^{IJ}_a(x)$ is a Hermitian $N\times N$ matrix.

Treating $\psi_I(x)$ and $\bar\psi_J(y)$ as (independent) Grassmann variables, and
using the Berezin integral $\int\cD[\psi]$,
we can compute correlation functions of the
fields from the formula
\begin{align}&
  \langle\psi_{I_1}(x_1)\cdots\psi_{I_n}(x_n)
  \bar\psi_{J_m}(y_m)\cdots \bar\psi_{J_1}(y_1)\rangle\\:= \nn
&  \frac{1}{Z}\int\e^{-\cS[\psi]}
  \psi_{I_1}(x_1)\cdots\psi_{I_n}(x_n)
  \bar\psi_{J_m}(y_m)\cdots \bar\psi_{J_1}(y_1)\cD[\psi]
  \end{align}
where 
\beq Z = \int\e^{-\cS[\psi]}\cD[\psi]  \eeq
is the normalization factor. Since these integrals tend to be infinite, some regularization is usually necessary to obtain physically meaningful results. Since the integrals are Gaussian (in the Grassmannian sense), all
correlation functions can be reduced to the correlation function for a pair of fields
\beq \langle\psi_I(x)\bar\psi_J(y)\rangle =- (D^{-1})_{IJ}(x,y;m,\Phi), \label{minus}\eeq
where $D^{-1}(x,y;m,\Phi)$ is the integral kernel of the inverse of
the operator $D(m,\Phi)$. Note the identity
\beq \langle \bar\psi_I(x) \Ga^a \psi_J(y)\rangle = \tr\big(\Ga^a  (D^{-1})_{JI}(y,x;m,\Phi)\big),\eeq
where the symbol $\tr$ denotes the trace over the spinor indices only,
and the minus sign from \eqref{minus} has disappeared, because  of the
anticommution relations of Grassmann variables.

\section{Axial anomaly}

In the introduction we introduced the vector current $\cJ^\mu(x)$,
and also
the axial currents $\cJ^{5}(x)$ and $\cJ^{5\mu}(x)$. Furthermore, we  defined
the anomaly $\cA(x)$.

We have (naively)
\beq \langle\pd_\mu \cJ^{5\mu}(x)\rangle = \sum_I \tr\big(\ga^5\ga^\mu \frac{\pd (D^{-1})_{II}(x,x;m,\Phi)}{\pd x^\mu} \big), \eeq
\beq \langle \cJ^{5}(x)\rangle = \sum_I \tr\big(\ga^5 (D^{-1})_{II}(x,x;m,\Phi) \big) ,\eeq
\beq \langle \bar\psi_I(x) \Ga^a \psi_J(x)\rangle = \tr\big(\Ga^a  (D^{-1})_{JI}(x,x;m,\Phi)\big).\eeq
These expressions however have a problem, because the integral kernel
$D^{-1}(x,y;m)$ is divergent for $y=x$.

To resolve this problem, we shall use Pauli-Villars regularization. Namely, we define regularized propagators (free and full) as
\beq (D_0^{-1})_\Lambda(x,y) = \sum_i \cC_i D_0^{-1}(x,y; M_i(\Lambda)),\eeq
\beq (D^{-1})_\Lambda(x,y;\Phi) = \sum_i \cC_i D^{-1}(x,y; M_i(\Lambda),\Phi),\eeq
where $\Lambda$ is a regularization parameter (eventually
$\Lambda\rarr +\infty$), masses $M_i(\Lambda)$ and coefficients
$\cC_i$ are chosen such that \[\cC_0=1,\quad M_0(\Lambda)=m,\quad \lim_{\Lambda\rarr +\infty} M_i(\Lambda) = +\infty\text{ for }i\neq 0,\] and $(D_0^{-1})_\Lambda(x,y)$ has no divergence for $y\rarr x$. These conditions mean that the integral
\beq (D_0^{-1})_\Lambda(x,x) = \int\frac{\rd^4 p}{(2\pi)^4} \sum_i \cC_i \frac{-\ri \pslash + M_i}{p^2+M_i^2} \eeq
  has to be finite. By \eqref{int-q1} and  \eqref{int-q1a},
  this is guaranteed by the conditions
\beq \sum_i \cC_i = 0, \qquad \sum_i \cC_i M_i = 0, \qquad \sum_i
\cC_i M_i^2 = 0, \qquad \sum_i \cC_i M_i^3 =
0. \label{CM-conditions}\eeq
A possible set solutions for these equations are
\beq i\in\{0,\dots n\},n\ge 4, \qquad \cC_i = (-1)^i\binom{n}{i}, \qquad M_i = m + i\Lambda .\label{CM-solution} \eeq
A proof that this is indeed a solution is given in Appendix \ref{Identity for Pauli-Villars regularization}. Other solutions can be created by linear combinations of these solutions. 

Using this regularized propagator we define
\beq \langle\pd_\mu J^{\mu}(x)\rangle_\Lambda = \sum_I \tr\big(\ga^\mu \frac{\pd (D^{-1})_{II,\Lambda}(x,x;\Phi)}{\pd x^\mu} \big), \eeq
\beq \langle\pd_\mu J^{5\mu}(x)\rangle_{\Lambda} = \sum_I \tr\big(\ga^5\ga^\mu \frac{\pd (D^{-1})_{II,\Lambda}(x,x;\Phi)}{\pd x^\mu} \big), \eeq
\beq \langle J^{5}(x)\rangle_{\Lambda} = \sum_I \tr\big(\ga^5 (D^{-1})_{II,\Lambda}(x,x;\Phi) \big) , \label{J5-regvev}\eeq
\beq \langle \bar\psi^I(x) \Ga^a \psi^J(x)\rangle_{\Lambda} = \tr\big(\Ga^a  (D^{-1})_{JI,\Lambda}(x,x;\Phi)\big), \label{Ja-regvev}\eeq
and
\begin{align} \cA_{\Lambda}(x;\Phi) &= \langle\pd_\mu J^{5\mu}(x)\rangle_{\Lambda} + 2m \langle J^{5}(x)\rangle_{\Lambda} + \nn\\
&\quad + \sum_{IJ}\langle \bar\psi_I(x) \big(\ga^5\Phi^{IJ}(x) + \Phi^{IJ}(x)\ga^5\big)\psi_J(x)\rangle_{\Lambda}.
\end{align}

We shall see that this regularization keeps the vector current conserved on the quantum level, i.e. 
\beq \lim_{\La\rarr\infty} \langle\pd_\mu J^{\mu}(x)\rangle_{\Lambda} = 0. \eeq
This relation is actualy satisfied even without the limit. We shall also see that the axial currents
in general produces infinite axial anomalies linear, quadratic, and
cubic in fields $\Phi$.
If the anomaly has a finite limit for $\Lambda\rarr\infty$, it's okay;
it's analogous to the standard textbook axial anomaly. However if in
this limit it diverges, we find it problematic.   
 We consider the disapearance of these
infinite anomalies a necessary condition for the consistency of the
quantization; for that to happen, fields $\Phi$ need to satisfy some
specific conditions, which we will derive in this paper.
Under these
conditions, we also
compute the resulting anomaly. Our result can be summarized as follows.

\bet\label{main}
Suppose that the following conditions are fullfilled:
\beq {\rm Tr}\,\la = 0, \label{finiteness-1}\eeq

\beq {\rm Tr}(\pd_\mu C^{\mu} +  2 \kappa\lambda +  2 B^{\mu\nu}\tilde B_{\mu\nu} ) = 0, \label{finiteness-2}\eeq

 \begin{align} &  {\rm Tr}\Big(- \ka^2\la + \ri \kappa[A^\mu, C_\mu] + 2\kappa B^{\mu\nu}\tilde B_{\mu\nu} - \ri[A^\mu, A^\nu] \tilde B_{\mu\nu} +\nn\\
&\qquad + \ri \tilde B_{\mu\nu} [C^\mu, C^\nu] + 2 B^{\mu\nu}B_{\mu\nu} \la + 2 C^\mu C_\mu \la + \la^3\Big) =  0.  \label{finiteness-3}\end{align} 
Then there exist the limit
\begin{align}
  \cA(x)&:= \lim_{\La\rarr\infty} \cA_\Lambda(x) ,
\end{align}
the relation \eqref{anomaly-definition0}  is satisfied and the renormalized anomaly is
\beq\cA = \cA^{(1)} + \cA^{(2)} + \cA^{(3)} + \cA^{(4)}, \eeq
where

\beq
\cA^{(1)} = \frac{1}{12\pi^2} {\rm Tr} \big(\pd^\mu\pd_\mu\pd_\nu C^{\nu}\big), \label{anomaly-1} \eeq

\begin{align}
\cA^{(2)} &= \frac{1}{12\pi^2} {\rm Tr} \Big(4 \kappa \pd^\mu\pd_\mu\lambda  + 6 \pd_\mu\kappa \pd^\mu \lambda  + \nn\\
&\quad + 3 \eps^{\mu\nu\al\be} \pd_\mu A_\nu\pd_\al A_\be + 2 A_\nu \pd^\mu\pd_\mu C^{\nu} - 2 C^{\nu} \pd^\mu\pd_\mu A_\nu    +\nn\\
&\quad + 4 \tilde B^{\rho\si}  \pd^\mu\pd_\mu  B_{\rho\si} + 8 \tilde B^{\rho\mu}  \pd_\mu \pd^\nu  B_{\rho\nu} - 8 B^{\rho\mu}  \pd_\mu \pd^\nu  \tilde B_{\rho\nu}  + \nn\\
&\quad +  6 g^{\mu\nu} \pd_\mu \tilde B^{\rho\si} \pd_\nu B_{\rho\si} + 4 \pd_\mu \tilde B^{\rho\mu} \pd^\nu B_{\rho\nu}- 4 \pd^\mu \tilde B^{\rho\nu} \pd_\nu B_{\rho\mu}  + \nn\\
&\quad - 12\, \pd^\mu B_{\mu\nu} \pd^\nu\lambda  + \eps^{\mu\al\nu\be} \pd_\mu C_\al \pd_\nu C_\be  \Big) ,\label{anomaly-2}
\end{align}

\begin{align}
\cA^{(3)} 
&= \frac{1}{12\pi^2}{\rm Tr}\bigg((\pd_\mu\ka)\Big( 10 \{\ka, C^\mu\}  + 4\ri  [\la, A^\mu] + 4 \{A_\nu, \tilde B^{\mu\nu}\} \Big)  + \nn\\
& +(\pd_\mu A_\nu)\Big(2 \ri g^{\mu\nu}[\ka,\la]   + 3 \ri \eps^{\mu\nu\al\be}[A_\al, A_\be]  - 3 \ri \eps^{\mu\nu\al\be} [C_\al, C_\be] + \nn\\
&\quad + 4 \{A^\mu, C^\nu\}  - 4\{A^\nu, C^\mu\}  - 2 g^{\mu\nu}\{A^\al, C_\al\}  + \nn\\
&\quad + 12 \{\ka, \tilde B^{\mu\nu}\} - 12 \{\la, B^{\mu\nu}\} - 12\ri g_{\al\be} (B^{\mu\al}  \tilde B^{\nu\be}  - B^{\nu\be} \tilde B^{\mu\al}) \Big)  + \nn\\
& + (\pd_\mu B_{\al\be})\Big(3 \eps^{\mu\nu\al\be}\{\ka, A_\nu\} - 4\ri g^{\mu\be} [\ka, C^\al]  + 4 \ri \eps^{\mu\nu\al\be}[\la, C_\nu]   + \nn\\
&\quad - 12\ri g^{\mu\be} [A_\nu, \tilde B^{\al\nu}] - 12 \ri  [A^\al, \tilde B^{\mu\be}]  + 4 \ri [A^\mu, \tilde B^{\al\be}] +\nn\\
&\quad + 16 g^{\mu\be} \{C_\nu, B^{\al\nu} \} + 8 \{C^\al, B^{\mu\be}\} - 4 \{C^\mu,B^{\al\be}\}  \Big) + \nn\\
& + (\pd_\mu C_\nu)\Big(10g^{\mu\nu}\ka^2 - 6 g^{\mu\nu}\la^2 + \nn\\
&\quad - 4 g^{\mu\nu} A^\al A_\al - 2 A^\mu A^\nu - 2 A^\nu A^\mu  + \nn\\
&\quad - 8 g^{\mu\nu} C^\al C_\al - 2 C^\mu C^\nu - 2 C^\nu C^\mu  + \nn\\
&\quad + 4 \ri [ \ka, B^{\mu\nu}] - 4 \ri [\la, \tilde B^{\mu\nu}]  - 4 g^{\mu\nu} B^{\rho\si}B_{\rho\si}  \Big) + \nn\\
& + (\pd_\mu\la)\Big(- 6\ri [\ka, A^\mu]  - 12 \ri[\tilde  B^{\mu\nu},C_\nu]  \Big) \bigg) , \label{anomaly-3}
\end{align}

\begin{align}
\cA^{(4)} &= \frac{1}{96\pi^2} {\rm Tr}\Big( -32 \ka^3\la  + 32\ri \ka^2[A^\mu, C_\mu] + 16 [A^\al, \ka][A_\al, \la]   + \nn\\
&\quad  + 48\ri \la^2[A^\mu, C_\mu]  + 48 C^\al C_\al (\ka\la+\la\ka) + 96\ka C^\mu \la C_\mu + \nn\\
&\quad  + 64 \ka^2 B^{\mu\nu}\tilde B_{\mu\nu} + 32\ka B^{\mu\nu}\ka\tilde B_{\mu\nu} + \nn\\
&\quad + 48 B^{\al\be}B_{\al\be} (\ka\la+\la\ka) + 64\ka B^{\mu\nu} \la B_{\mu\nu}+\nn\\
&\quad - 192 \la^2 B^{\mu\nu}\tilde B_{\mu\nu}  - 96\la B^{\mu\nu}\la \tilde B_{\mu\nu} \nn\\
&\quad - 48\ri (\tilde B_{\mu\nu} \ka + \ka \tilde B_{\mu\nu}) [A^\mu, A^\nu]  + 48\ri (B_{\mu\nu} \la + \la B_{\mu\nu})[A^\mu, A^\nu]  + \nn\\
&\quad  + 16\ri (\tilde B_{\mu\nu} \ka + \ka \tilde B_{\mu\nu} ) [C^\mu, C^\nu] + 128\ri \tilde B_{\mu\nu} C^\mu \ka C^\nu +\nn\\
&\quad - 16\ri (B_{\mu\nu}\la + \la B_{\mu\nu})[C^\mu, C^\nu]- 96\ri B_{\mu\nu} C^\mu \la C^\nu +\nn\\
&\quad + 32(B_{\mu\nu}\ka +  \ka B_{\mu\nu})[A^\mu, C^\nu] + 32 (\tilde B_{\mu\nu} \la + \la  \tilde B_{\mu\nu})[A^\mu, C^\nu] + \nn\\
&\quad - 64 B_{\mu\nu}  (A^\mu \ka C^\nu + C^\nu \ka A^\mu) - 64 \tilde B_{\mu\nu} (\la A^\mu C^\nu + C^\nu A^\mu \la) +\nn\\
&\quad + 64 \tilde B_{\mu\nu} (A^\mu \la C^\nu + C^\nu\la A^\mu) + 64 \tilde B_{\mu\nu} (\la C^\nu A^\mu + A^\mu C^\nu  \la) + \nn \\
&\quad - 256\ri g_{\mu\nu} \ka B^{\mu\al}\tilde B_{\al\be} B^{\be\nu} - 512\ri g_{\mu\nu} \la B^{\mu\al} B_{\al\be} B^{\be\nu} + \nn\\
&\quad - 6\eps^{\mu\nu\al\be}[A_\mu, A_\nu][A_\al, A_\be] + 2\eps^{\mu\nu\al\be}[C_\mu, C_\nu][C_\al, C_\be] + \nn\\
&\quad  - 4\eps^{\mu\nu\al\be}[A_\mu, A_\nu][C_\al, C_\be] + 8\eps^{\mu\nu\al\be}[A_\mu, C_\nu][A_\al, C_\be] + \nn\\
&\quad - 16\ri [A_\mu, A_\nu][A^\mu, C^\nu] + 32\ri A^\mu A_\mu [A^\nu, C_\nu] + \nn\\
&\quad + 32 A^\al A_\al B^{\mu\nu}\tilde B_{\mu\nu} - 32 A^\al B^{\mu\nu} A_\al \tilde B_{\mu\nu} - 128  g_{\al\be} B^{\mu\al} \tilde B^{\nu\be} [A_\mu, A_\nu] +\nn\\
&\quad -96 C^\al C_\al B^{\mu\nu}\tilde B_{\mu\nu} + 32 C^\al B^{\mu\nu} C_\al \tilde B_{\mu\nu} +64 \ri  g_{\al\be} B^{\mu\al} \tilde B^{\nu\be} [C_\mu, C_\nu] + \nn\\
&\quad + 192\ri ( A_\mu B_{\nu\al} C^\nu B^{\mu\al} - A_\mu B^{\mu\al} C^\nu B_{\nu\al}) + 64\ri B^{\al\be}B_{\al\be} [A^\mu, C_\mu]  + \nn\\
&\quad + 64\ri(A_\mu C^\nu B_{\nu\al} B^{\mu\al} - A_\mu B^{\mu\al} B_{\nu\al} C^\nu ) + \nn\\
&\quad - 64  B^{\al\be}B_{\al\be} B^{\mu\nu}\tilde B_{\mu\nu} + 64  [B^{\al\mu}, B_{\be\mu}]   [B_{\al\nu} , \tilde B^{\be\nu} ] \Big) . \label{anomaly-4}
\end{align}

\eet

Among many terms in the above theorem, we can find the terms that
reproduce the standard axial anomaly of Yang-Mills field:

\begin{align}
& \cA(x)_{\kappa,B,C,\lambda=0} =\nn\\
&= \frac{1}{4\pi^2}\eps^{\mu\nu\al\be} \Tr(\pd_\mu A_\nu \pd_\al A_\be) + \frac{\ri}{2\pi^2}\eps^{\mu\nu\al\be} \Tr(\pd_\mu A_\nu A_\al A_\be) + \nn\\
&\quad + \frac{1}{16\pi^2}\eps^{\mu\nu\al\be} \Tr([A_\mu, A_\nu][ A_\al, A_\be]) = \nn\\
& = \frac{1}{16\pi^2} \eps^{\mu\nu\al\be} \Tr(A_{\mu\nu} A_{\al\be}),\end{align}
where $A_{\mu\nu} = \pd_\mu A_\nu - \pd_\nu A_\mu + \ri[A_\mu,A_\nu]$.

Another part that we can extract from the full result are the terms that depend only on scalar fields $\kappa$ and $\lambda$, that is when $A_\mu=0$, $B_{\mu\nu}=0$, $C_\mu=0$:

\beq \cA(x)
_{A,B,C=0} = \frac{1}{12\pi^2} \Tr(4\ka \pd^\mu \pd_\mu\la + 6\pd^\mu \ka \pd_\mu\la - 4\ka^3\la), \eeq 
with the conditions for the vanishing of divergent part being
\begin{align} \tr(\la)&=0, \\
 \tr(\ka\la) &= 0,\\
\tr( - \ka^2\la + \la^3)&=0.\end{align}

One more special case would be when the configuration of external fields has $A_\mu=0$, $B_{\mu\nu}=0$, $\la=0$:

\begin{align} \cA(x)_{A,B,\la=0} &= \frac{1}{12\pi^2} \Tr\big(\eps^{\mu\nu\rho\si}\pd_\mu C_\nu \pd_\rho C_\si +  10 \pd_\mu(\ka ^2 C^\mu)  \big) = \nn\\
&= \frac{1}{12\pi^2} \pd_\mu \Tr\big(\eps^{\mu\nu\rho\si} C_\nu \pd_\rho C_\si +  10 \ka ^2 C^\mu  \big),\end{align}
with the only condition for the vanishing of divergent part being
\beq \tr(\pd_\mu C^\mu) = 0.\eeq

\section{First steps of proof}

The remainder of the paper is devoted to a proof of the above theorem.


From the definition of $D^{-1}(x,y; M,\Phi)$ we have:
\beq \big(\gamma^\mu\frac{\pd}{\pd x^\mu} + M + \Phi(x)\big)D^{-1}(x,y; M,\Phi) = \de^4(x-y) ,\eeq
\beq D^{-1}(x,y; M,\Phi) \big(-\gamma^\mu\frac{\overleftarrow{\pd}}{\pd y^\mu} + M + \Phi(y)\big) = \de^4(x-y) ,\eeq
therefore
\begin{align} & \big(\gamma^\mu\pd_\mu + m +\Phi(x)\big)(D^{-1})_{\Lambda}(x,y;\Phi)  \nn\\
&= \sum_{i} \cC_i \big(\gamma^\mu\pd_\mu + m + \Phi(x)\big) D^{-1}(x,y; M_i(\Lambda),\Phi) \nn\\
&= \sum_{iJ} \cC_i \Big(\big(m  - M_i(\Lambda)\big) D^{-1} (x,y; M_i(\Lambda),\Phi) + \de^4(x-y)\Big)  \nn\\
&= \sum_{i} \cC_i \big(m- M_i(\Lambda)\big) D^{-1}(x,y; M_i(\Lambda),\Phi) .
\end{align}

Let's first check the conservation of the vector current:
\begin{align} &\langle\pd_\mu J^{\mu}(x)\rangle_{\Lambda} \nn \\
=& \lim_{y\rarr x}  \Tr\,\tr\Big( \ga^\mu\frac{\pd}{\pd x^\mu}  (D^{-1})_{\Lambda}(x,y;\Phi) + (D^{-1})_{\Lambda}(x,y;\Phi) \ga^\mu\frac{\overleftarrow{\pd}}{\pd y^\mu} \Big) \nn \\
=& \lim_{y\rarr x} \bigg( \Tr\,\tr\Big(  \sum_{i}\cC_i\big( - M_i(\Lambda) - \Phi(x) \big) D^{-1}(x,y;M_i(\Lambda),\Phi) +  \de^4(x-y)\Big)  \nn\\
&\quad\qquad + \Tr\,\tr\Big(  \sum_{i}\cC_i D^{-1}(x,y;M_i(\Lambda),\Phi) \big(M_i(\Lambda)  + \Phi(y)\big) -  \de^4(x-y)  \Big) \bigg) \nn \\
=& 0 .\end{align}

We can express the regularized anomaly as
\begin{align} &\cA_{\Lambda}(x;\Phi) \nn  \\
=& \lim_{y\rarr x}  \Tr\,\tr\Big(\ga^5\big( \ga^\mu\frac{\pd}{\pd x^\mu} + m + \Phi(x)\big) (D^{-1})_{\Lambda}(x,y;\Phi)\Big)  \nn\\
&\quad\qquad + \tr\Big(\ga^5 (D^{-1})_{\Lambda}(x,y;\Phi) \big(-\ga^\mu\frac{\overleftarrow{\pd}}{\pd y^\mu} + m) + \Phi(y)\big)\Big) \nn \\
 =& \lim_{y\rarr x} \sum_{i} 2 \cC_i \big(m - M_i(\Lambda)\big)
    \Tr\,\tr\big(\ga^5 D^{-1}(x,y; M_i(\Lambda),\Phi) \big)
    . \label{reganom}
\end{align}

Let us interrupt for a moment the proof of Theorem \ref{main} to
comment on whether the anomalies we compute are gauge invariant.
Unfortunately,
while operator $D(m,\Phi)$ satisfies relation \eqref{gauge-transformation}, and thus we have also
\beq U^{-1}D^{-1}(M_i(\La),\Phi)U = D\big(M_i(\La), U^{-1}\Phi U  + \ga^\mu U^{-1}(\pd_\mu U)\big) \eeq
\beq U^{-1}(D^{-1})_\La(\Phi)U = (D^{-1})_\La\big(U^{-1}\Phi U  + \ga^\mu U^{-1}(\pd_\mu U)\big) \eeq
no similar relation will be satisfied by $\cA_{\Lambda}(x;\Phi)$ as we
defined it.


In fact, by  \eqref{reganom} we have
\begin{align} & \cA_{\Lambda}(x;U^{-1}\Phi U  + \ga^\mu U^{-1}(\pd_\mu U)) = \nn \\
& = \lim_{y\rarr x} \sum_{i} 2 \cC_i \big(m - M_i(\Lambda)\big) \Tr\,\tr\big(\ga^5 D^{-1}(x,y; M_i(\Lambda),U^{-1}\Phi U  + \ga^\mu U^{-1}(\pd_\mu U)) \big) = \nn\\
&= \lim_{y\rarr x} \sum_{i} 2 \cC_i \big(m - M_i(\Lambda)\big) \Tr\,\tr\big(\ga^5 U^{-1}(x) D^{-1}(x,y; M_i(\Lambda),\Phi) U(y) \big) = \nn\\
&= \lim_{y\rarr x} \sum_{i} 2 \cC_i \big(m - M_i(\Lambda)\big) \Tr\,\tr\big(\ga^5 U(y)U^{-1}(x) D^{-1}(x,y; M_i(\Lambda),\Phi)  \big)
\end{align}
which means that
\begin{align} & \cA_{\Lambda}(x;U^{-1}\Phi U  + \ga^\mu U^{-1}(\pd_\mu U)) - \cA_{\Lambda}(x;\Phi)= \nn \\
&= \lim_{y\rarr x} \Tr\,\tr\big(\ga^5 (U(y)U^{-1}(x) - 1) \sum_{i} 2 \cC_i \big(m - M_i(\Lambda)\big)  D^{-1}(x,y; M_i(\Lambda),\Phi)  \big) \end{align}
While combination $\sum_i \cC_i D^{-1}(x,y; M_i(\Lambda),\Phi)$ has no divergence for $y\rarr x$ (with appropriately chosen coefficient $\cC_i$), it's not necessarily true for \newline $\sum_i \cC_i M_i D^{-1}(x,y; M_i(\Lambda),\Phi)$. Because of that divergence, the limit in the expresion above does not vanish, and it contains terms dependent of $U$.

Let us go back to the proof of Theorem \ref{main}.
Expanding $D^{-1}(M) = (D_0(M) + \Phi)^{-1}$ into a series with regard to the powers of $\Phi$:
\begin{align}\notag D^{-1}(M) = &D_0^{-1}(M)  - D_0^{-1}(M) \Phi D_0^{-1}(M) \\& +
D_0^{-1}(M) \Phi D_0^{-1}(M)\Phi D_0^{-1}(M)+ \dots \end{align}
we  obtain the expansion of $\cA_{\text{reg}}(x;\Lambda)$ into a series:
\beq \cA_{\Lambda}(x)  = \cA_{\Lambda}^{(0)}(x) + \cA_{\Lambda}^{(1)}(x) + \cA_{\Lambda}^{(2)}(x) + \dots\eeq
in which $\cA_{\Lambda}^{(n)}$ is a homogenous polynomial of $n$-th order in fields $\Phi$.

\section{Zeroth-order term}

Using \eqref{int-q1} and  \eqref{int-q1a}, we obtain
\begin{align} \cA_{\Lambda}^{(0)}(x) &= \lim_{y\rarr x} \sum_{i} 2 \cC_i \big(m - M_i\big) \Tr\,\tr\big(\ga^5 D_0^{-1}(x,y; M_i)   \big) = \nn \\
&= N \int\frac{\rd^4 p}{(2\pi)^4}  \sum_{i} 2 \cC_i (m - M_i) \tr\big(\ga^5 \frac{-\ri \pslash  + M_i}{p^2+M_i^2} \big) = \nn \\
&= 0. \end{align}

\section{First-order term}

\begin{align} &\cA_{\Lambda}^{(1)}(x) = \nn\\
&= \lim_{y\rarr x} (-1) \sum_{i} 2 \cC_i (m - M_i) \Tr\,\tr \big(\ga^5 \int\rd^4z D_0^{-1}(x,z; M_i) \Phi(z) D_0^{-1}(z,y; M_i)\big) = \nn \\
&= -\int\frac{\rd^4 p}{(2\pi)^4} \int\frac{\rd^4 q}{(2\pi)^4} \re^{\ri p x} \sum_i 2 \cC_i \big(m - M_i\big) \times \nn\\
&\quad \times \Tr\,\tr\bigg( \frac{ (-\ri \qslash + M_i)\ga^5 (-\ri (\qslash+\pslash) + M_i)  }{((q+p)^2+M_i^2)(q^2+M_i^2)} \Phi(p)  \bigg) .
\end{align}

To calculate this expression, we'll use the identity (\ref{pr-ga5-pr}) and then (\ref{int-q1}), (\ref{int-q2}), (\ref{int-q3}). We get


\begin{align} & \cA_{\Lambda}^{(1)}(x) = \nn\\
&= -\int\frac{\rd^4 p}{(2\pi)^4} \re^{\ri p x} \bigg(\frac{1}{(4\pi)^2 } \sum_i 2\cC_i (m-M_i) M_i^2 \log( M_i^2) \Tr\,\tr(\ga^5\Phi(p) ) + \nn\\
&\qquad  + \frac{-\ri p_\mu}{(4\pi)^2} \Big( \sum_i 2\cC_i (m-M_i) M_i \log (M_i)^2 + \frac13 p^2 + \cO(\La^{-1}) \Big) \Tr\,\tr(\ga^\mu \ga^5 \Phi(p) ) + \nn\\
&\qquad  + \frac{p^2}{(4\pi)^2} \Big( \frac12  \sum_i 2\cC_i (m-M_i)   \log (M_i^2) + \cO(\La^{-1})\Big)\Tr\,\tr(\ga^5 \Phi(p) ) \bigg)  = \nn \\
&= \frac{-1}{2\pi^2}\Tr\big(\la(x)\big)  \sum_{i} \cC_i \big(m - M_i\big) M_i^2 \log (M_i^2) + \nn\\
 &\qquad + \frac{-1}{2\pi^2}\Tr\big(\pd_\mu C^{\mu}(x) \big) \sum_{i}\cC_i (m-M_i) M_i  \log (M_i^2)  + \nn\\
&\qquad + \frac{1}{4\pi^2}\Tr\big(\pd^\mu\pd_\mu\la(x)\big)  \sum_{i}\cC_i (m-M_i) \log (M_i^2)  + \nn\\
 &\qquad + \frac{1}{12\pi^2}\Tr\big(\pd^\mu\pd_\mu\pd_\nu C^{\nu}(x)\big)   + \cO(\Lambda^{-1}).  \end{align}

We see that some terms are divergent with $\Lambda$. For now, we'll just keep them in mind, as similar divergent terms may arise in terms of higher order in $\Phi$. The finite term will be denoted as $\cA^{(1)}$, and matches the formula given in  (\ref{anomaly-1}).

\section{Second-order term}

\begin{align} &\cA_{\Lambda}^{(2)}(x) = \nn\\
&= \lim_{y\rarr x} \sum_{i} 2 \cC_i (m - M_i) \int\rd^4z_1 \int\rd^4z_2 \, \nn \\
&\qquad \times \Tr\,\tr\big(\ga^5 D_0^{-1}(x,z_1; M_i) \Phi(z_1) D_0^{-1}(z_1,z_2; M_i) \Phi(z_2) D_0^{-1}(z_2,y; M_i) \big) = \nn \\
=& \iint\frac{\rd^4p\rd^4k}{(2\pi)^8} \re^{\ri (p+k) x} \int\frac{\rd^4q}{(2\pi)^4}\sum_{i} \cC_i 2(m-M_i)  \times\nn\\&\quad\times  \Tr\,\tr\big(\Phi(p)D_0^{-1}(q;M_i)\Phi(k)D_0^{-1}(q-k;M_i)\ga^5 D_0^{-1}(q+p;M_i)\big) . \end{align}

Using identity (\ref{pr-ga5-pr}) and then (\ref{int-q2}-\ref{int-q7}), we get

\begin{align}
&\cA_{\Lambda}^{(2)}(x) = \nn\\
=& \iint\frac{\rd^4p\rd^4k}{(2\pi)^8} \re^{\ri (p+k) x} \times\nn\\
&\quad\times \bigg(\frac{\ri }{(4\pi)^2} \sum_{i}  \cC_i 2(m-M_i) \log (M_i^2) \times \nn\\
&\quad\qquad \times \Big(-\frac12\Tr\,\tr\big( \Phi(p) \pslash  \Phi(k)  \ga^5 \big) + \frac{1}{4} \Tr\,\tr\big( \Phi(p)\ga_\mu \Phi(k) \ga^\mu (\pslash+\kslash)\ga^5\big)\Big) + \nn\\
&\qquad + \frac{-1}{(4\pi)^2} \Big(\sum_{i} \cC_i 2(m-M_i)M_i \log (M_i^2)\Big) \Tr\,\tr\big( \Phi(p) \Phi(k)  \ga^5 \big)  + \nn\\
&\qquad + \frac{1}{3(4\pi)^2} p^2 \Tr\,\tr\big( \Phi(p) \Phi(k)  \ga^5 \big)   + \nn\\
&\qquad + \frac{1}{3(4\pi)^2}\Tr\,\tr\big( \Phi(p) \Phi(k) (-2\kslash-\pslash) (\pslash+\kslash)\ga^5\big)   + \nn\\
&\qquad + \frac{1}{3(4\pi)^2}\Tr\,\tr\big( \Phi(p)(\kslash-\pslash) \Phi(k) (\pslash+\kslash)\ga^5\big) + \nn\\
&\qquad + \frac{- \ri m}{2(4\pi)^2} \Tr\,\tr\big( \Phi(p) \Phi(k) (\pslash+\kslash)\ga^5\big) \bigg) + \cO(\Lambda^{-1}) .\label{Aren2-traces}
\end{align}

Calculating the traces we obtain

\begin{align}
&\cA_{\Lambda}^{(2)}(x) = \nn\\
=& \frac{-1}{\pi^2} \Tr\Big( \kappa(x)\lambda(x) + B^{\mu\nu}(x)\tilde B_{\mu\nu}(x)  \Big) \sum_{i} \cC_i (m-M_i)M_i \log (M_i^2)   + \nn \\
& + \cA^{(2)} + \cO(\Lambda^{-1}),
\end{align}
where $\cA^{(2)}$, the finite part, is given by eq. (\ref{anomaly-3}).

Again, we can see a probematic term divergent when $\Lambda \rarr \infty$. It has the same kind of divergence as one of the terms that appeared in the linear order, and they will need to be considered together, leading to the condition (\ref{finiteness-2}). Among the many finite terms there's one proportional to  $\Tr(\eps^{\mu\nu\al\be} \pd_\mu A_\nu(x)\pd_\al A_\be(x))$, which leads to the standard, well-known axial anomaly in the presence of vector field.

\section{Third-order term}

\begin{align} &\cA_{\Lambda}^{(3)}(x) = \nn\\
&= \lim_{y\rarr x} (-1)\sum_{i} 2 \cC_i (m - M_i) \int\rd^4z_1 \int\rd^4z_2 \int\rd^4z_3 \times \nn \\
&\qquad \times \Tr\big(\ga^5 D_0^{-1}(x,z_1; M_i) \Phi(z_1) D_0^{-1}(z_1,z_2; M_i) \Phi(z_2) \times \nn \\
&\qquad \qquad \times D_0^{-1}(z_2,z_3; M_i) \Phi(z_3) D_0^{-1}(z_3,y; M_i)\big) = \nn \\
&= -\iiint\frac{\rd^4p_1\rd^4p_2\rd^4p_3}{(2\pi)^{12}} \re^{\ri (p_1+p_2+p_3) x} \times\nn\\
&\quad\times \int\frac{\rd^4q}{(2\pi)^4}\sum_{i\neq 0}  \cC_i 2(m-M_i) \tr\big(\Phi(p_1) D_0^{-1}(q;M_i)\Phi(p_2) \times\nn\\ &\qquad\qquad\times  D_0^{-1}(q-p_2;M_i) \Phi(p_3)  D_0^{-1}(q-p_2-p_3;M_i)\ga^5 D_0^{-1}(q+p_1;M_i) \big).  \end{align}

Using identity (\ref{pr-ga5-pr}) we get

\begin{align} & D_0^{-1}(q-p_2-p_3;M_i)\ga^5 D_0^{-1}(q+p_1;M_i) = \nn\\
&= \frac{1}{(q+p_1)^2+M_i^2}\Big(1 + D_0^{-1}(q-p_2-p_3;M_i)\ri(\pslash_1+\pslash_2+\pslash_3) \Big)\ga^5 . \end{align}

The second term in this expression leads to the integrals which are
convergent even without the sum $\sum_{i}  \cC_i 2(m-M_i)$. The
leading term in those integrals is proportional to $1/M_i$, which
makes them $\cO(\Lambda^{-1})$ dla $i\neq 0$. That means that only
first term from the expression above can lead to a term that are
divergent in the limit $\Lambda\rarr\infty$. The second term can still
give a finite contribution, thanks to the factor $(m-M_i)$, and simple
dimensional analysis tells us that in  the part of integral over $q$ that contains this second term we can omit all momenta $p_i$ except for the factor $\ri(\pslash_1+\pslash_2+\pslash_2)$---all terms coming from these omitted momenta would be at least of order $(m-M_i)/M_i^2$, which means they would vanish in the limit $\Lambda\rarr\infty$.

\begin{align} &\cA_{\Lambda}^{(3)}(x) = \nn\\
&= -\iiint\frac{\rd^4p_1\rd^4p_2\rd^4p_3}{(2\pi)^{12}} \re^{\ri (p_1+p_2+p_3) x}   \times\nn\\
&\quad\times \Big(\int\frac{\rd^4q}{(2\pi)^4}\sum_{i}  \cC_i 2(m-M_i) \frac{1}{(q+p_1)^2+M_i^2} \times\nn\\ &\qquad\qquad\times   \tr\big(\Phi(p_1) D_0^{-1}(q;M_i)\Phi(p_2)D_0^{-1}(q-p_2;M_i) \Phi(p_3)\ga^5 \big) + \nn\\
&\qquad+ \int\frac{\rd^4q}{(2\pi)^4}\sum_{i}  \cC_i 2(m-M_i) \frac{\ri (p_1+p_2+p_3)_\mu}{(q+p_1)^2+M_i^2} \times\nn\\ &\qquad\quad\times   \tr\big(\Phi(p_1) D_0^{-1}(q;M_i) \Phi(p_2) D_0^{-1}(q-p_2;M_i)\times\nn\\
&\qquad\qquad \times \Phi(p_3) D_0^{-1}(q-p_2-p_3;M_i) \gamma^\mu \ga^5 \big)  \Big) = \nn\\
&= -\iiint\frac{\rd^4p_1\rd^4p_2\rd^4p_3}{(2\pi)^{12}} \re^{\ri (p_1+p_2+p_3) x} \int\frac{\rd^4q}{(2\pi)^4}\sum_{i}  \cC_i 2(m-M_i)   \times\nn\\
&\quad\times \bigg(\frac{-q_\nu (q-p_2)_\mu }{((q+p_1)^2+M_i^2)(q^2+M_i^2)((q-p_2)^2+M_i^2)}  \times\nn\\
 &\qquad\qquad\times   \tr\big(\Phi(p_1) \ga^\nu \Phi(p_2)\ga^\mu \Phi(p_3)\ga^5 \big) + \nn\\
&\qquad+ \frac{-\ri q_\mu M_i }{((q+p_1)^2+M_i^2)(q^2+M_i^2)((q-p_2)^2+M_i^2)}  \times\nn\\
 &\qquad\qquad\times   \tr\big(\Phi(p_1) \ga^\mu \Phi(p_2) \Phi(p_3)\ga^5 \big) + \nn\\
&\qquad+ \frac{-\ri (q-p_2)_\mu M_i }{((q+p_1)^2+M_i^2)(q^2+M_i^2)((q-p_2)^2+M_i^2)}  \times\nn\\
 &\qquad\qquad\times   \tr\big(\Phi(p_1) \Phi(p_2) \ga^\mu \Phi(p_3)\ga^5 \big) + \nn\\
&\qquad+\frac{M_i^2}{((q+p_1)^2+M_i^2)(q^2+M_i^2)((q-p_2)^2+M_i^2)}  \times\nn\\
 &\qquad\qquad\times   \tr\big(\Phi(p_1) \Phi(p_2) \Phi(p_3)\ga^5 \big) + \nn\\
&\qquad+\ri (p_1+p_2+p_3)_\al \Big( \frac{-q_\mu q_\nu M_i}{(q^2+M_i^2)^4} + (\text{terms with $p_j$})  \Big) \times\nn\\ 
&\qquad\quad\times   \tr\Big(\Phi(p_1) \big(\ga^\mu\Phi(p_2) \Phi(p_3)\ga^\nu + \ga^\mu\Phi(p_2)\ga^\nu \Phi(p_3)+\Phi(p_2)\ga^\mu \Phi(p_3) \ga^\nu\big) \gamma^\al \ga^5  \Big) + \nn\\
&\qquad+\ri (p_1+p_2+p_3)_\al\Big( \frac{M_i^3}{(q^2+M_i^2)^4} + (\text{terms with $p_j$})  \Big) \times\nn\\ 
&\qquad\quad\times   \tr\big(\Phi(p_1) \Phi(p_2) \Phi(p_3) \gamma^\al \ga^5  \big)\bigg) .
\end{align}

Using identities (\ref{int-q7}), (\ref{int-q4}), (\ref{int-q5}), (\ref{int-q6}), (\ref{int-q10}) and (\ref{int-q11}), we get

\begin{align} &\cA_{\Lambda}^{(3)}(x) = \nn\\
&= -\iiint\frac{\rd^4p_1\rd^4p_2\rd^4p_3}{(2\pi)^{12}} \re^{\ri (p_1+p_2+p_3) x} \sum_{i\neq 0}  \cC_i 2(m-M_i)   \times\nn\\
&\quad\times \bigg(\Big( \frac{g_{\mu\nu}}{(4\pi)^2} \log (M_i^2) + \cO(\La^{-2}) \Big)    \Tr\big(\Phi(p_1) \ga^\mu \Phi(p_2)\ga^\nu \Phi(p_3)\ga^5 \big) + \nn\\
&\qquad+ \Big(\frac{-\ri}{6(4\pi)^2} \frac{(p_2-p_3)_\mu}{M_i} + \cO(\La^{-3}) \Big)    \Tr\big(\Phi(p_1) \ga^\mu \Phi(p_2) \Phi(p_3)\ga^5 \big) + \nn\\
&\qquad+ \Big(\frac{-\ri}{6(4\pi)^2} \frac{(-2p_2-p_3)_\mu}{M_i} + \cO(\La^{-3}) \Big)   \Tr\big(\Phi(p_1) \Phi(p_2) \ga^\mu \Phi(p_3)\ga^5 \big) + \nn\\
&\qquad+\Big(\frac{1}{2(4\pi)^2}  + \cO(\La^{-2}) \Big)  \Tr\big(\Phi(p_1) \Phi(p_2) \Phi(p_3)\ga^5 \big) + \nn\\
&\qquad+\ri (p_1+p_2+p_3)_\al \Big(\frac{-g_{\mu\nu}}{12(4\pi)^2 M_i}  + \cO(\La^{-3}) \Big)   \times\nn\\ 
&\qquad\quad\times   \Tr\Big(\Phi(p_1) \big(\ga^\mu\Phi(p_2) \Phi(p_3)\ga^\nu + \ga^\mu\Phi(p_2)\ga^\nu \Phi(p_3)+\Phi(p_2)\ga^\mu \Phi(p_3) \ga^\nu\big) \gamma^\al \ga^5  \Big) + \nn\\
&\qquad+\ri (p_1+p_2+p_3)_\al \Big(\frac{1}{6(4\pi)^2 M_i}  + \cO(\La^{-3}) \Big)  \times\nn\\ 
&\qquad\quad\times   \Tr\big(\Phi(p_1) \Phi(p_2) \Phi(p_3) \gamma^\al \ga^5  \big)\bigg) = \nn \\
&= -\frac{1}{4(4\pi)^2} \sum_{i} \cC_i 2(m-M_i) \log(M_i^2) \Tr\big(\Phi \ga^\mu \Phi \ga_\mu \Phi \ga^5 \big) + \nn\\
&\quad+ \frac{-1}{3(4\pi)^2} \Tr\big(\Phi (\ga^\mu \pd_\mu\Phi  \Phi - \ga^\mu \Phi  \pd_\mu\Phi-2 \pd_\mu\Phi   \ga^\mu \Phi - \Phi \ga^\mu\pd_\mu\Phi)\ga^5 \big) + \nn\\
&\quad + \frac{1}{6(4\pi)^2} \pd_\mu\Tr\big(\Phi(2\Phi^2 - \ga^\nu\Phi^2 \ga_\nu - \ga^\nu\Phi\ga_\nu \Phi  - \Phi\ga^\nu \Phi\ga_\nu) \ga^\mu \ga^5 \big)  +\nn\\
&\quad+ \cO(\Lambda^{-1}), \end{align}
where in the last expression $\Phi$ means $\Phi(x)$ and $\pd_\mu$ means $\frac{\pd}{\pd x^\mu}$. After calculating the traces we get

\begin{align} &\cA_{\Lambda}^{(3)}(x) = \nn\\
=& \Big(\sum_{i} \cC_i 2(m-M_i) \log(M_i^2) \Big) \times \nn\\
&\quad\times \frac{-1}{4\pi^2} \Tr\Big(\ka^2\la - \ri \kappa[A^\mu, C_\mu] -2\kappa B^{\mu\nu}\tilde B_{\mu\nu} + \ri[A^\mu, A^\nu] \tilde B_{\mu\nu} +\nn\\
&\qquad -\ri \tilde B_{\mu\nu} [C^\mu, C^\nu]- B^{\mu\nu}B_{\mu\nu} \la - 2 C^\mu C_\mu \la - \la^3\Big) + \nn\\
&+ \cA^{(3)}  + \cO(\La^{-1}),\end{align}
where $\cA^{(3)}$, the finite part, is given by eq. (\ref{anomaly-3}), and the divergent part contributes to the condition (\ref{finiteness-3}).

\section{Fourth-order term}

\begin{align} &\cA_{\Lambda}^{(4)}(x) = \nn\\
&= \lim_{y\rarr x} \sum_{i} 2 \cC_i (m - M_i) \int\rd^4z_1 \int\rd^4z_2 \int\rd^4z_3 \int\rd^4z_4\, \nn \\
&\qquad \times \Tr\,\tr\big(\ga^5 D_0^{-1}(x,z_1; M_i) \Phi(z_1) D_0^{-1}(z_1,z_2; M_i) \Phi(z_2) \nn \\
&\qquad \qquad \times D_0^{-1}(z_2,z_3; M_i) \Phi(z_3) D_0^{-1}(z_3,z_4; M_i)  \Phi(z_4) D_0^{-1}(z_4,y; M_i)\big) = \nn \\
&= \iiiint\frac{\rd^4p_1\rd^4p_2\rd^4p_3\rd^4p_4}{(2\pi)^{16}} \re^{\ri (p_1+p_2+p_3+p_4) x} \int\frac{\rd^4q}{(2\pi)^4}\sum_{i}  \cC_i 2(m-M_i)  \times\nn\\
&\quad\times \Tr\,\tr\Big(D_0^{-1}(q+p_1+p_2+p_3+p_4;M_i)\Phi(p_1) D_0^{-1}(q+p_2+p_3+p_4;M_i) \times\nn\\
 &\qquad\times \Phi(p_2) D_0^{-1}(q+p_3+p_4;M_i) \Phi(p_3)  D_0^{-1}(q+p_4;M_i) \Phi(p_4) D_0^{-1}(q;M_i)\ga^5  \Big) .\end{align}

Using identity (\ref{pr-ga5-pr}) we get

\begin{align} & D_0^{-1}(q;M_i)\ga^5 D_0^{-1}(q+p_1+p_2+p_3+p_4;M_i) = \nn\\
&= \Big(\frac{1}{(q+p_1+p_2+p_3+p_4)^2+M_i^2} + \frac{(-\ri \qslash + M_i)\ri(\pslash_1+\pslash_2+\pslash_3+\pslash_4)}{(q^2+M_i^2)((q+p_1+p_2+p_3+p_4)^2+M_i^2)} \Big)\ga^5 . \end{align}

Using dimensional analysis we can see that the second term will only produce contributions that vanish in the limit $\La\rarr\infty$. The first term may produce a finite contribution, but  this contribution will be independent of $p_i$.

\begin{align} &\cA_{\Lambda}^{(4)}(x) = \nn\\
&= \iiiint\frac{\rd^4p_1\rd^4p_2\rd^4p_3\rd^4p_4}{(2\pi)^{16}} \re^{\ri (p_1+p_2+p_3+p_4) x} \int\frac{\rd^4q}{(2\pi)^4}\sum_{i}  \cC_i 2(m-M_i) \frac{1}{q^2+M_i^2}  \times\nn\\
 &\qquad\times \Tr\, \tr\Big(\Phi(p_1)D_0^{-1}(q;M_i)\Phi(p_2)D_0^{-1}(q;M_i) \Phi(p_3)  D_0^{-1}(q;M_i) \Phi(p_4) \ga^5  \Big) + \cO(\Lambda^{-1}) = \nn \\
&= \int\frac{\rd^4q}{(2\pi)^4}\sum_{i}  \cC_i 2(m-M_i) \bigg(\frac{M_i^3}{(q^2+M_i^2)^4} \Tr\,\tr\big(\Phi^4 \ga^5  \big) + \nn\\
&\quad + \frac{-q_\mu q_\nu M_i}{(q^2+M_i^2)^4} \Tr\,\tr\Big(\Phi (\ga^\mu \Phi \ga^\nu \Phi + \ga^\mu \Phi^2 \ga^\nu  + \Phi \ga^\mu \Phi \ga^\nu )\Phi \ga^5  \Big) \bigg) + \nn\\
&\quad + \cO(\Lambda^{-1}) \nn =\\
&= \sum_{i\neq 0}  \cC_i 2(m-M_i) \bigg(\frac{1}{6(4\pi)^2 M_i} \Tr\,\tr\big(\Phi^4 \ga^5  \big) + \nn\\
&\quad + \frac{-g_{\mu\nu}}{12(4\pi)^2 M_i} \Tr\,\tr\Big(\Phi (\ga^\mu \Phi \ga^\nu \Phi + \ga^\mu \Phi^2 \ga^\nu  + \Phi \ga^\mu \Phi \ga^\nu )\Phi \ga^5  \Big) \bigg) + \nn\\
&\quad + \cO(\Lambda^{-1}) = \nn \\
&= \frac{1}{6(4\pi)^2} \Tr\,\tr\Big(\Phi (2\Phi^2 - \ga^\mu \Phi \ga_\mu \Phi - \ga^\mu \Phi^2 \ga_\mu  - \Phi \ga^\mu \Phi \ga_\mu )\Phi \ga^5  \Big) +\nn\\
&\quad + \cO(\Lambda^{-1})  = \nn\\
&= \frac{1}{6(4\pi)^2}  \Tr\Big( -32 \ka^3\la  + 32\ri \ka^2[A^\mu, C_\mu] - 16A^\al A_\al (\ka\la+\la\ka)  + 32 A^\al \ka A_\al \la   + \nn\\
&\quad  + 48\ri \la^2[A^\mu, C_\mu]  + 48 C^\al C_\al (\ka\la+\la\ka) + 96\ka C^\mu \ka C_\mu + \nn\\
&\quad  + 64 \ka^2 B^{\mu\nu}\tilde B_{\mu\nu} + 32\ka B^{\mu\nu}\ka\tilde B_{\mu\nu} + \nn\\
&\quad + 24 B^{\al\be}B_{\al\be} (\ka\la+\la\ka) + 64\ka B^{\mu\nu} \la B_{\mu\nu}+\nn\\
&\quad - 192 \la^2 B^{\mu\nu}\tilde B_{\mu\nu}  - 96\la B^{\mu\nu}\la \tilde B_{\mu\nu} \nn\\
&\quad - 48\ri (\tilde B_{\mu\nu} \ka + \ka \tilde B_{\mu\nu}) [A^\mu, A^\nu]  + 48\ri (B_{\mu\nu} \la + \la B_{\mu\nu})[A^\mu, A^\nu]  + \nn\\
&\quad  + 16\ri (\tilde B_{\mu\nu} \ka + \ka \tilde B_{\mu\nu} ) [C^\mu, C^\nu] + 128\ri \tilde B_{\mu\nu} C^\mu \ka C^\nu +\nn\\
&\quad - 16\ri (B_{\mu\nu}\la + \la B_{\mu\nu})[C^\mu, C^\nu]- 96\ri B_{\mu\nu} C^\mu \la C^\nu +\nn\\
&\quad + 32(B_{\mu\nu}\ka +  \ka B_{\mu\nu})[A^\mu, C^\nu] + 32 (\tilde B_{\mu\nu} \la + \la  \tilde B_{\mu\nu})[A^\mu, C^\nu] + \nn\\
&\quad - 64 B_{\mu\nu}  (A^\mu \ka C^\nu + C^\nu \ka A^\mu) + \nn\\
&\quad - 64 \tilde B_{\mu\nu} (\la A^\mu C^\nu + C^\nu A^\mu \la) +\nn\\
&\quad + 64 \tilde B_{\mu\nu} (A^\mu \la C^\nu + C^\nu\la A^\mu) + \nn\\
&\quad + 64 \tilde B_{\mu\nu} (\la C^\nu A^\mu + A^\mu C^\nu  \la) + \nn \\
&\quad - 256\ri g_{\mu\nu} \ka B^{\mu\al}\tilde B_{\al\be} B^{\be\nu} - 512\ri g_{\mu\nu} \la B^{\mu\al} B_{\al\be} B^{\be\nu} + \nn\\
&\quad  + 32\ri A^\al A_\al [A^\mu, C_\mu] + \nn\\
&\quad + 32 A^\al A_\al B^{\mu\nu}\tilde B_{\mu\nu} - 32 A^\al B^{\mu\nu} A_\al \tilde B_{\mu\nu} + \nn\\
&\quad +40\ri B^{\al\be}B_{\al\be} [A^\mu, C_\mu] - 16  B^{\al\be}B_{\al\be} B^{\mu\nu}\tilde B_{\mu\nu} + \nn\\
&\quad -96 C^\al C_\al B^{\mu\nu}\tilde B_{\mu\nu} + 32 C^\al B^{\mu\nu} C_\al \tilde B_{\mu\nu} + \nn\\
&\quad + 192\ri ( A_\mu B_{\nu\al} C^\nu B^{\mu\al} -A_\mu B^{\mu\al} C^\nu B_{\nu\al}) + \nn\\ 
&\quad + 64\ri(A_\mu C^\nu B_{\nu\al} B^{\mu\al} - A_\mu B^{\mu\al} B_{\nu\al} C^\nu ) + \nn\\
&\quad - 6\eps^{\mu\nu\al\be}[A_\mu, A_\nu][A_\al, A_\be] + \nn\\
&\quad - 16\ri [A_\mu, A_\nu][A^\mu, C^\nu] + \nn\\
&\quad - 4\eps^{\mu\nu\al\be}[A_\mu, A_\nu][C_\al, C_\be] + \nn\\
&\quad + 8\eps^{\mu\nu\al\be}[A_\mu, C_\nu][A_\al, C_\be] + \nn\\
&\quad + 2\eps^{\mu\nu\al\be}[C_\mu, C_\nu][C_\al, C_\be] + \nn\\
&\quad - 128  g_{\al\be} B^{\mu\al} \tilde B^{\nu\be} [A_\mu, A_\nu] +\nn\\
&\quad +64 \ri  g_{\al\be} B^{\mu\al} \tilde B^{\nu\be} [C_\mu, C_\nu] + \nn\\
&\quad + 64  [B^{\al\mu}, B_{\be\mu}]   [B_{\al\nu} , \tilde B^{\be\nu} ] \Big) + \cO(\Lambda^{-1})  ,\end{align}
where fields without an argument mean fields at a point $x$.

\section{Summary}

Higher order integrals include at least six propagators $D_0^{-1}(p,M)$, so even with the factor $(m-M_i)$ they will not produce terms divergent or finite in the limit $\Lambda\rarr\infty$. The only divergent terms come from the terms up to the 3-rd order in $\Phi$, and the finite terms are up to the 4-th order in $\Phi$, which we calculated above. Writing all of them together we have

\begin{align} &\cA_{\Lambda}(x) = \nn\\
&= \Big(\sum_{i} \cC_i \big(m - M_i\big) M_i^2 \log (M_i^2) \Big) \frac{-1}{2\pi^2} \Tr(\la)  + \nn\\
&\quad + \Big(\sum_{i}\cC_i (m-M_i) M_i  \log (M_i^2)\Big)  \times \nn\\
&\qquad \times \Big( \frac{-1}{2\pi^2}\Tr(\pd_\mu C^{\mu})    + \frac{-1}{\pi^2}\Tr ( \kappa\lambda +   B_{\mu\nu}\tilde B_{\mu\nu} ) \Big)  + \nn\\
&\quad + \Big(\sum_{i} \cC_i 2(m-M_i) \log(M_i^2) \Big) \bigg( \sum_{I} \frac{1}{4\pi^2}\Tr(\Box\la)  + \nn\\
&\quad  + \frac{-1}{4\pi^2} \Tr\Big(\ka^2\la - \ri \kappa[A^\mu, C_\mu] -2\kappa B^{\mu\nu}\tilde B_{\mu\nu} + \ri[A^\mu, A^\nu] \tilde B_{\mu\nu} +\nn\\
&\qquad -\ri \tilde B_{\mu\nu} [C^\mu, C^\nu]- 2 B^{\mu\nu}B_{\mu\nu} \la - 2 C^\mu C_\mu \la - \la^3\Big) \bigg) + \cO(\Lambda^0).\end{align}

For the divergent terms to vanish, we need then

\beq \tr (\la ) = 0 ,\eeq

\beq \tr\Big(\pd_\mu C^{\mu} + 2 \kappa\lambda +  2 B^{\mu\nu}\tilde B_{\mu\nu} \Big) = 0, \eeq
\begin{align} &  \tr\Big(\pd^\mu\pd_\mu\la  - \ka^2\la + \ri \kappa[A^\mu, C_\mu] + 2\kappa B^{\mu\nu}\tilde B_{\mu\nu} - \ri[A^\mu, A^\nu] \tilde B_{\mu\nu} +\nn\\
&\qquad + \ri \tilde B_{\mu\nu} [C^\mu, C^\nu] + 2 B^{\mu\nu}B_{\mu\nu} \la + 2 C^\mu C_\mu \la + \la^3\Big) =  0.\end{align}

We require these conditions to be satsfied everyhwere, in the whole space. From the first condition it follows then that $\tr(\pd^\mu\pd_\mu\la) = 0$ and this term can be omitted from the third condition, leaving us with:

\begin{align} &  \tr\Big(- \ka^2\la + \ri \kappa[A^\mu, C_\mu] + 2\kappa B^{\mu\nu}\tilde B_{\mu\nu} - \ri[A^\mu, A^\nu] \tilde B_{\mu\nu} +\nn\\
&\qquad + \ri \tilde B_{\mu\nu} [C^\mu, C^\nu] + 2 B^{\mu\nu}B_{\mu\nu} \la + 2 C^\mu C_\mu \la + \la^3\Big) =  0.\end{align} 

Optionally, we can also use the second condition to obtain the identity

\begin{align} \cA^{(1)} &= \frac{1}{12\pi^2} {\rm Tr} \big(\pd^\mu\pd_\mu\pd_\nu C^{\nu}\big) = \nn\\
&= \frac{1}{12\pi^2} \pd^\mu\pd_\mu {\rm Tr} \big( -2 \kappa\lambda -  2 B^{\rho\si}\tilde B_{\rho\si} \big). \end{align}

This way, we can get rid of the term $\cA^{(1)}$ completely, adding additional terms to $\cA^{(2)}$ instead. Whether it is worth it, may be situational.

If these conditions are fullfilled, the regularized anomaly has a finite limit for $\La\rarr\infty$, which contains terms from the first to the fourth order in fields:

\begin{align}& \cA(x) = \lim_{\La\rarr\infty} \cA_{\Lambda}(x) = \cA^{(1)}(x) + \cA^{(2)}(x) + \cA^{(3)}(x) + \cA^{(4)}(x).
\end{align}

\appendix
\section{Formula library}

In this appendix we collect various formulas useful in the proof of
Theorem \ref{main}. We start with the elementary identity
\beq \frac{-\ri \pslash_2 + M_i}{p_2^2+M_i^2} \ga^5 \frac{-\ri \pslash_1 + M_i}{p_1^2+M_i^2}  = \Big(\frac{1}{p_1^2+M_i^2} + \frac{(-\ri \pslash_2 + M_i)\ri(\pslash_1-\pslash_2)}{(p_2^2+M_i^2)(p_1^2+M_i^2)} \Big)\ga^5  \label{pr-ga5-pr}\eeq

The following family of identities is very useful when we consider the
so-called Feynman parameters:
\beq \int_0^\infty \rd\rho \sum_i A_i \re^{-\rho B_i} = \sum_i \frac{A_i}{B_i}\eeq

Assuming $\sum_i A_i = 0$:
\beq \int_0^\infty \rd\rho\,\rho^{-1} \sum_i A_i \re^{-\rho B_i} = - \sum_i A_i \log B_i\eeq

Assuming $\sum_i A_i = 0$, $\sum_i A_i B_i =0$:
\beq \int_0^\infty \rd\rho\,\rho^{-2} \sum_i A_i \re^{-\rho B_i} = \sum_i A_i B_i \log B_i\label{qwqw}\eeq

Let us formulate the remaining identities as lemmas. Note that all
integrands in the following integrals are in $L^1$.
\begin{lemma}
Assuming $\sum_i A_i = 0$, $\sum_i A_i M_i^2 =0$:
\begin{align} \int \frac{\rd^4 q}{(2\pi)^4} \sum_i A_i \frac{1}{q^2+M_i^2},
&= \frac{1}{(4\pi)^2 } \sum_i A_i M_i^2 \log( M_i^2) ,\label{int-q1}\\
\int \frac{\rd^4 q}{(2\pi)^4} \sum_i A_i \frac{q^\mu}{q^2+M_i^2}&=0 .
\label{int-q1a}\end{align}
\end{lemma}

\proof The integrability of the integrand of
\eqref{int-q1} follows from
\begin{align}\notag
\sum_i A_i \frac{1}{q^2+M_i^2}=&
\sum_i A_i
\Big(\frac{1}{q^2+M_i^2}-\frac{1}{q^2+m^2}+\frac{M_i^2-m^2}{(q^2+m^2)^2}\Big)\\
=&\sum_i A_i \frac{(m^2-M_i^2)^2}{(q^2+M_i^2)(q^2+m^2)^2}.
\end{align}
To obtain \eqref{int-q1} we first introduce the so-called Feynman
representation, then we evaluate the Gaussian integral, and finally
we apply \eqref{qwqw}:
\begin{align}\notag
\int \frac{\rd^4 q}{(2\pi)^4} \sum_i A_i 
\frac{1}{q^2+M_i^2}=&\int_0^\infty\rd\rho\int 
\frac{\rd^4 q}{(2\pi)^4} \sum_i A_i 
\e^{-(q^2+M_i^2)\rho}\\=\;
\int_0^\infty\rd\rho
\frac{1}{(4\pi)^2} \sum_i \frac{A_i }{\rho^2}
\e^{-M_i^2\rho}=&
\frac{1}{(4\pi)^2} \sum_i A_i M_i^2\log(M_i^2).
\end{align}
A proof of \eqref{int-q1a} is left to the reader.
\qed

We also skip the proofs of the following identities:
\begin{lemma}
Assuming $\sum_i A_i = 0$:
\begin{align} &\int \frac{\rd^4 q}{(2\pi)^4} \sum_i A_i \frac{1}{\big((q+p)^2+M_i^2\big)(q^2+M_i^2)} = \nn\\
&= \frac{-1}{(4\pi)^2} \int_0^1\rd\be_1\int_0^1\rd\be_2\,\de(1-\be_1-\be_2) \sum_i A_i \log (\be_1\be_2p^2 + M_i^2), \label{int-q2}
 \end{align}
\begin{align} &\int \frac{\rd^4 q}{(2\pi)^4} \sum_i A_i \frac{q_\mu}{\big((q+p)^2+M_i^2\big)(q^2+M_i^2)} = \nn\\
&= \frac{p_\mu}{(4\pi)^2} \int_0^1\rd\be_1\int_0^1\rd\be_2\,\de(1-\be_1-\be_2) \sum_i A_i \be_1 \log (\be_1\be_2p^2 + M_i^2), \label{int-q3}
 \end{align}
\begin{align} & \int\frac{\rd^4q}{(2\pi)^4} \sum_i A_i \frac{q_\nu (q-k)_\mu }{(q^2+M_i^2)((q-k)^2+M_i^2)((q+p)^2+M_i^2)} =\nn\\
&= \frac{1}{(4\pi)^2} \int_0^1\rd\be_1\int_0^1\rd\be_2\int_0^1\rd\be_3\, \de(1-\be_1-\be_2-\be_3)  \times \nn\\
&\quad\qquad \times \Big( \sum_i A_i\frac{(\be_2 k - \be_3 p)_\nu (\be_3 (-k-p) -\be_1 k)_\mu}{\be_1\be_2 k^2 + \be_1\be_3 p^2 + \be_2\be_3 (k+p)^2 + M_i^2}  + \nn\\
&\quad\qquad\qquad -\frac{g_{\mu\nu}}{2} \sum_i A_i\log\big(\be_1\be_2 k^2 + \be_1\be_3 p^2 + \be_2\be_3 (k+p)^2 + M_i^2\big) \Big) ,\label{int-q7}
\end{align}
\begin{align}
& \int\frac{\rd^4q}{(2\pi)^4}\sum_{i} A_i \frac{q_\mu q_\nu}{(q^2+M_i^2)^3} =
 \frac{- g_{\mu\nu}}{4(4\pi)^2} \sum_{i} A_i \log (M_i^2) .\label{int-q9}
\end{align}
\end{lemma}

\begin{lemma}
\begin{align} & \int\frac{\rd^4q}{(2\pi)^4} \frac{1}{(q^2+M_i^2)((q-k)^2+M_i^2)((q+p)^2+M_i^2)} =\nn\\
&= \frac{1}{(4\pi)^2} \int_0^1\rd\be_1\int_0^1\rd\be_2\int_0^1\rd\be_3\, \de(1-\be_1-\be_2-\be_3) \times \nn\\
&\quad\qquad \times \frac{1}{\be_1\be_2 k^2 + \be_1\be_3 p^2 + \be_2\be_3 (k+p)^2 + M_i^2} ,\label{int-q4}
\end{align}

\begin{align} & \int\frac{\rd^4q}{(2\pi)^4} \frac{q_\mu}{(q^2+M_i^2)((q-k)^2+M_i^2)((q+p)^2+M_i^2)} =\nn\\
&= \frac{1}{(4\pi)^2} \int_0^1\rd\be_1\int_0^1\rd\be_2\int_0^1\rd\be_3\, \de(1-\be_1-\be_2-\be_3) \times \nn\\
&\quad\qquad \times \frac{(\be_2 k -\be_3 p)_\mu }{\be_1\be_2 k^2 + \be_1\be_3 p^2 + \be_2\be_3 (k+p)^2 + M_i^2} ,\label{int-q5}
\end{align}

\begin{align} & \int\frac{\rd^4q}{(2\pi)^4} \frac{(q-k)_\mu}{(q^2+M_i^2)((q-k)^2+M_i^2)((q+p)^2+M_i^2)} =\nn\\
&= \frac{1}{(4\pi)^2} \int_0^1\rd\be_1\int_0^1\rd\be_2\int_0^1\rd\be_3\, \de(1-\be_1-\be_2-\be_3) \times \nn\\
&\quad\qquad \times \frac{(\be_3 (-k-p) -\be_1 k)_\mu }{\be_1\be_2 k^2 + \be_1\be_3 p^2 + \be_2\be_3 (k+p)^2 + M_i^2} ,\label{int-q6}
\end{align}

\begin{align}
& \int\frac{\rd^4q}{(2\pi)^4}  \frac{1}{(q^2+M_i^2)^3} = 
 \frac{1}{2(4\pi)^2M_i^2} ,\label{int-q8}
\end{align} 

\begin{align}
& \int\frac{\rd^4q}{(2\pi)^4}  \frac{1}{(q^2+M_i^2)^4} = 
 \frac{1}{6(4\pi)^2M_i^4} ,\label{int-q10}
\end{align}

\begin{align}
& \int\frac{\rd^4q}{(2\pi)^4}  \frac{q_\mu q_\nu}{(q^2+M_i^2)^4} = 
 \frac{g_{\mu\nu}}{12(4\pi)^2M_i^2} .\label{int-q11}
\end{align}
\end{lemma}

\section{Identity for Pauli-Villars regularization}
\label{Identity for Pauli-Villars regularization}

A proof that (\ref{CM-solution}) is a solution of (\ref{CM-conditions}): \\
Using (\ref{CM-solution}) we have
\begin{align} \sum_i \cC_i M_i^k &= \sum_{i=0}^n (-1)^i \binom{n}{i} (m + i\Lambda)^k =\nn\\
&= \sum_{p=0}^k \binom{k}{p} m^{k-p}\Lambda^p \sum_{i=0}^n (-1)^i \binom{n}{i} i^p . \end{align}
We have then
\begin{align} & \sum_{i=0}^n (-1)^i \binom{n}{i} i^p =\nn \\
& = \Big( \sum_{i=0}^n (-1)^i \binom{n}{i} i^p x^i\Big)\Big|_{x=1}  = \nn \\
& = \Big(\big(x \frac{\rd}{\rd x}\big)^p\sum_{i=0}^n (-1)^i \binom{n}{i} x^i\Big)\Big|_{x=1} = \nn \\
& = \Big(\big(x \frac{\rd}{\rd x}\big)^p (1-x)^n \Big)\Big|_{x=1}.\end{align}
Since $x=1$ is a n-th order zero of function $(1-x)^n$, then even
after applying the operator $x \frac{\rd}{\rd x}$ less than $n$ times,
it will still be a zero of the resulting function; therefore we get 
\beq \sum_{i=0}^n (-1)^i \binom{n}{i} i^p = 0,  \qquad \text{for }p<n ;\eeq
\beq \sum_i \cC_i M_i^k = 0, \qquad \text{for }k<n.\eeq

\paragraph{Acknowledgement.} 
The support of the National Science Center of Poland under the 
    grant UMO-2019/35/B/ST1/01651 is acknowledged.

\end{document}